\documentstyle[11pt]{article}
\input{epsf}

\def\1{\mbox{I\hspace{-.15em}1}}

\def\b{\begin{equation}}
\def\e{\end{equation}}
\def\bee{\begin{enumerate}}
\def\eee{\end{enumerate}}

\title{Experimental structure of gravitational wave detection by bounded \\cold electronic plasma in a long pipe}
\author{K.Mehdizadeh$^{1}$, O. Jalili$^{2}$\thanks{E-Mail: omid\_jalili@yahoo.com },  and M.V.
Takook$^{3}$\thanks{E-Mail: takook@razi.ac.ir}}
\date{\today}

\begin{document}
  \maketitle {\centerline {\it $^1$ Department of physics,
   Islamic Azad University-Ayatollah Amoli branch,} \centerline{\it P.O.BOX
678, Amol, IRAN,}  \centerline {\it $^2$ Department of physics,
   Islamic Azad University-Noor branch} \centerline{\it P.O.BOX
46415/444, Noor, Mazandaran, IRAN,} \centerline {\it $^3$ Department of Physics, Razi University, Kermanshah, Iran }

\begin{abstract}

In the previous paper, we introduced a new method of gravitational waves (GW) detection\cite{jalili}. In our proposal, we replaced usual Weber's metallic bar with a cold electronic plasma. We obtained a nonhomogenous differential equation for tangential electric field, $E_\phi$, that on it GW is known as nonhomogenous term. In this paper we estimate, the dimension of pipe, the electron density and some other associated parameters for obtaining the best detection.

\end{abstract}

\vspace{0.5cm} {\it Proposed PACS numbers}: 04.62.+v, 98.80.Cq,
12.10.Dm \vspace{0.5cm}

\section{Introduction}

One of the two usual method to detect GW is the measurement of the length change in metallic bar \cite{Weber}. In the presence of GW the length of the bar will oscillate, then by measurement of the bars's length, we can indirectly find the amplitude of GW. In our previous paper \cite{jalili}, we used electronic column instead of metallic bars. Figure-1 schematically shows or proposal.

\begin{figure}[tb]
  \centerline{\epsfxsize=4.5in \epsfysize=1.5in{\epsffile{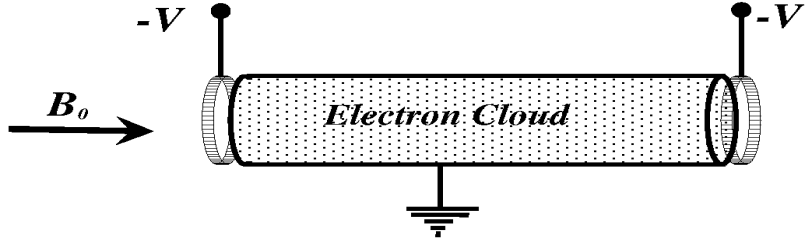}}}
    \caption{Enclosed cold electronic plasma in a long pipe. Two negative potential rings at the end of the pipe create a potential barrier.}\label{11}
\end{figure}

The enclosed electrons rotate about external longitudinal magnetic field, simultaneously go back and forth along the cylinder. At the end of the cylinder the electrons will be reflected and go back into the cylinder. After the GW collide with this bar, some electrons will get enough energy and they will pass through the potential barriers. In this paper, we determine some physical parameters of the apparatus by considering the dynamics of the electrons.
In section 2, we discuss the dynamics of the enclosed electrons and with applying  the two conditions, stability and the confinement of the electrons in the pipe, we get a permissible region for the electrons in the $n-B$ plane. In section 3, by considering electromagnetic field due to GW, we estimate the strength of such perturbative electromagnetic field and then get strength of other fields such as density and velocity fields. Then the order of electrons's  perturbative  energy will be obtained. Having these results, we adjust the potential barrier height to confine the electron in the cylinder. At the end, we could be able to draw the perturbative energy in term of the colliding GW strength. Using these information we can infer at what strength of GW, our apparatus will be able to detect GW.

\section{Steady state dynamics of the confined electrons within the pipe, without the GW }

The confined electrons within the pipe rotate with the two following frequency \cite{jalili}:
\b\omega _\pm=\frac{\omega _c}{2}[1\pm(1-
\frac{2\omega_p^2}{\omega_c^2})^\frac{1}{2}] ,\e where $
\omega_c=\frac{eB_0}{mc}$  is the cyclotron frequency of
pseudo-neutral plasma, $\omega_p=\sqrt{\frac{4\pi n e^2}{m}}$ the plasma frequency, $m$ the electron mass, n the electron density, c the speed of light, $e$ the electron charge and $B_0$ the magnetic field. For obtaining a real $\omega _\pm$, we must have $\frac{2\omega_p^2}{\omega_c^2}\leq1$. This is our first condition. This condition physically means as follows. An electron inside the pipe  suffers two forces; repulsion force due to nonneutral nature of our electronic plasma and the magnetic centripetal force. Inequality of $\frac{2\omega_p^2}{\omega_c^2}\leq1$ states that  magnetic restoring forces (as measured by $\omega_c^2$) must overcome electrostatic repulsive forces (as measured by$\omega_p^2$), for radial confinement. If $\frac{2\omega_p^2}{\omega_c^2}>1$ beam expands radially, which is not an equilibrium state \cite{david}.
Replacing $\omega_p$ and $\omega_c$ in the condition  $\frac{2\omega_p^2}{\omega_c^2}\leq1$, we obtain the following condition in the $n-B$ plane:
\b n< \frac{1}{8\pi m n c^2}B_0 ^2 .\e
Fig-2 shows this curve, in fact our desired region is the lower region of this curve.

\begin{figure}[tb]
  \centerline{\epsfxsize=4.5in \epsfysize=3.5in{\epsffile{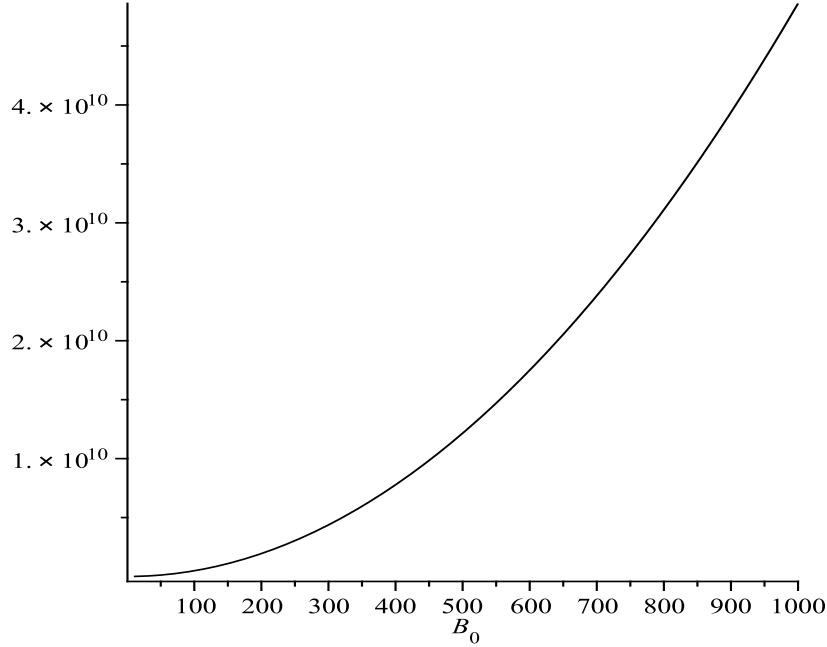}}}
    \caption{Stability condition, $\frac{2\omega_p^2}{\omega_c^2}\leq1$, is drown in $n-B$ plane. Permissible region is the lower region of this curve.}\label{fgt}
\end{figure}

Up to now, we obtained the first constrain on our parameters.
Another condition must be satisfied is the confinement condition, that means the electrons energy must be less than potential barrier height. Then:
\b E=\frac{1}{2}mv^2=\frac{1}{2}mv_\varphi^2=\frac{1}{2}m(r\omega_\pm)^2 \leq e\triangle\psi,\e
where $v_\varphi$ is the $\varphi$ component of the velocity, $\omega_\pm$ the angular frequency of the electrons, $\bigtriangleup\psi$ the potential difference between rings and the body of cylinder. Replacing Eq. (1) in Eq. (3) and doing some algebraic calculation, we obtain:

\b  \frac{m \omega_c^2}{8\pi e^2} \{1-[\sqrt{\frac{2e\triangle\varphi}{m}}(\frac{2}{r\omega_c})-1]^2\}\leq n . \e
Fig-3 show this Eq. in the $n-B$ plane.

\begin{figure}[tb]
  \centerline{\epsfxsize=4.5in \epsfysize=3.5in{\epsffile{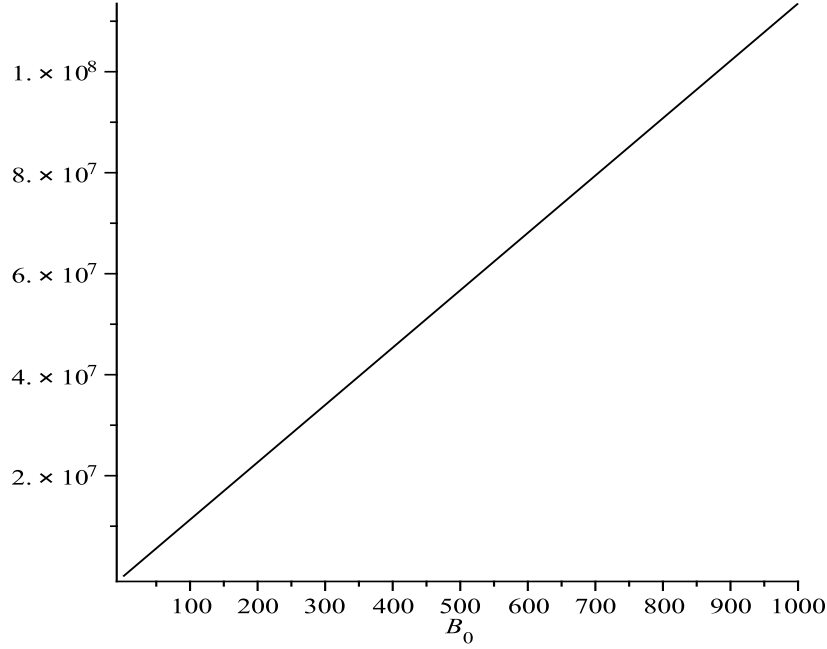}}}
    \caption{Confinement condition, $\frac{m \omega_c^2}{8\pi e^2} \{1-[\sqrt{\frac{2e\triangle\varphi}{m}}(\frac{2}{r\omega_c})-1]^2\}\leq n $, is drown in $n-B$ plane. Permissible region is the upper region of this curve.}\label{fgt}
\end{figure}

So far we found that for confinement and stability conditions, solution region in the $n-B$ plane is the area between the two curves in Fig 2 and Fig 3. Now, we try to determine other parameters of the system. Therefore, we must estimate the order of the explicit field in term of the order of the GW field.

\section{The order of perturbative electrons energy due to GW}

As GW collide with our system, rotational motion of electron about longitudinal axis, will be perturbed and a special electromagnetic field will be created. In our previous paper \cite{jalili}, we consider such a special mode. We showed in that paper, the perturbed tangential electric field $E_\phi$, satisfying the following differential equation:

\b
(\frac{i\omega}{c}+f')E_\phi+A''\frac{\partial}{\partial
r}(rE_\phi)+a'M(r)\frac{\partial}{\partial
r}(rE_\phi)-a'N(r)E_\phi $$ $$ =\frac{-c}{i\omega+\sigma
c}\frac{\partial}{\partial r}(\frac{1}{r}\frac{\partial}{\partial
r}(rE_\phi))+ \frac{i kc}{i\omega+\sigma
c}\frac{\partial}{\partial
r}(\frac{M(r)}{r}\frac{\partial}{\partial r}(rE_\phi))
-\frac{ikc}{i\omega+\sigma c}\frac{\partial}{\partial
r}(\frac{N(r)}{r}E_\phi)-\frac{b'c}{i\omega+\sigma
c}\frac{\partial}{\partial r}(rE_\phi)$$ $$
+\frac{ikb'c}{i\omega+\sigma c}M(r) \frac{\partial}{\partial
r}(rE_\phi) -\frac{ikb'c}{i\omega+\sigma
c}N(r)E_\phi-\frac{S''c}{i\omega+\sigma c}\frac{\partial}{\partial
r}(r\frac{\partial}{\partial r}(rE_\phi))$$ $$
+(g'+2G'')r\sigma-\frac{ikc\sigma}{i\omega+\sigma
c}\frac{\partial}{\partial
r}(\frac{L(r)}{r})-\frac{ikb'c\sigma}{i\omega+\sigma c}L(r),\e
where $a$, $b$, $f$, $g$, $A$, $D$, $S$, $G$,  M(r), N(r) and L(r) are defined as follows:
  \b   a=\frac{-e\omega_r/m}{(k\omega_r-\omega)^2-\omega_r^2}
,\;\;b=\frac{-e\omega^2_r/mc}{(k\omega_r-\omega)^2-\omega_r^2}=\frac{\omega_r}{c}a
,\;\;
f=\frac{i(k\omega_r-\omega)/m}{(k\omega_r-\omega)^2-\omega_r^2}
,\;\;
g=\frac{i(k\omega_r-\omega)/e}{(k\omega_r-\omega)^2-\omega_r^2}\omega_r,
$$ $$ A=\frac{-e}{m\omega_r}+\frac{(k\omega_r-\omega)^2}{\omega_rm[(k\omega_r-\omega)^2-\omega_r^2]} ,\;\;
 D=\frac{i(k\omega_r-\omega)e/m}{(k\omega_r-\omega)^2-\omega_r^2}
 ,\;\;
S=\frac{ie\omega_r(k\omega_r-\omega)/mc}{(k\omega_r-\omega)^2-\omega_r^2},$$
$$
G=[\frac{-e}{m\omega_r}+\frac{(k\omega_r-\omega)^2}{\omega_rm[(k\omega_r-\omega)^2
 -\omega_r^2]} ]\frac{m\omega_r}{e}
 =A\frac{m\omega_r}{e},\e

\b
M(r)=(\frac{ik}{r}-S'r)(\frac{c}{r(i\omega+\sigma c)})/R(r) .\;\;
N(r)=A'/R(r) ,\;\; L(r)=G'r/R(r) ,$$ $$
R(r)\equiv(\frac{-i\omega}{c}+D')+(\frac{ik}{r}-S'r)(\frac{c}{r(i\omega+\sigma
c)})+\frac{ik'^2c}{\omega} . \e

The primed and double primed quantity is defined as follow
\b  A'=\frac{-4\pi e n_0}{c}A ,\;\; a'=(\frac{-4\pi e
n_0}{c})(1+\frac{k\omega_r}{(\omega-k\omega_r)})a ,\;\;
A''=\frac{-4\pi e \omega_r n_0}{ic(\omega-k\omega_r)}A .\e

In the above expressions, $k$ is the wave number along the $z$ axis, $k'$ is the wave number along the $\phi$ axis and $\sigma$ is the amplitude of the shear tensor $\cite{Elli2}$.
Having $E_\phi(r)$, $E_r(r)$ component can be obtained from the following Equation \cite{jalili}:

\b E_{r}(r)=M(r)\frac{\partial}{\partial
r}(rE_{\phi(r)})-N(r)E_{\phi}(r)-L(r)\sigma.\e
The magnetic field will be derived from Faraday law as follow: \b
B_{\phi}=\frac{c k'}{\omega}E_{r},\e \b
B_{z}=\frac{c}{r(i\omega+\sigma c)}(\frac{\partial}{\partial
r}(rE_{\phi})-ikE_{r}).\e

Perturbative velocity field can be obtained from the continuum equation and force balance equation as follow \cite{jalili}:

\b v_\phi=aE_r+brB_z+fE_\phi+gr\sigma ,\;\;v_r=AE_\phi+DE_r+SrB_z+Gr\sigma. \e

Finally the perturbative electron energy can be obtained as follow:

\b E=\triangle(\frac{1}{2}mv)^2=m(v_{0\phi}\triangle v_\phi+v_{0r}\triangle v_r)=mv_{o\phi}\triangle v_\phi, \e

where $v_{0\phi}=r\omega$ is the electron velocity before GW presence and $\triangle v_\phi$ is the perturbative velocity.

Now, we try to estimate the order of these fields. Assuming that the variation of the fields due to radius is not so strong, then by Eq (5) we have:

\b o(E_\phi)=\{(g'+G'')r\sigma-\frac{i k c \sigma}{i \omega+\sigma c}\frac{\partial}{\partial r}(\frac{L}{r})-\frac{i k b' c \sigma L}{i \omega+\sigma c}\}/\{ (\frac{i \omega}{c}+f')+A''+a'M-a'N$$ $$-\frac{c}{(i \omega+\sigma c)r^2}-
\frac{i k c}{i \omega+\sigma c}\frac{\partial}{\partial r}(\frac{N-M}{r})+\frac{b'c-i k b' c M+i k b'c N+S''c}{i \omega+\sigma c}\}.\e
Thus the order of other fields can be obtained from Eq. (6), Eq. (7) and Eq. (8):
\b o(E_r)=(M-N)o(E_\phi)-L\sigma,\e
\b o(B_\phi)=\frac{c k'}{\omega}o(E_r),\e
\b o(B_z)=\frac{c}{r(i\omega+\sigma c)}(o(E_{\phi})-i k o(E_r)).\e
Thus presence of GW will create the peturbative field that finally will cause a perturbative electron's energy that can be obtained from Eq. (13).

\section{ Discussion and conclusions }

As we pointed out, We decided to optimize our proposed apparatus, namely to find  the best sensitivity of the system to GW. The parameters that must be specified were: diameter of the pipe, length of the pipe, potential difference between the rings and the pipe, the electrons density and axial magnetic field (one can find angular frequency of the electrons by these parameters). Then, we try to determine these parameters. First, we work in very low temperature to get rid of the unwanted electromagnetic field namely noise. This situation has a additional advantage: with low temperature, we can sure the Maxwellian distribution of the electrons velocity will sharply concentrate  about its average value, and thus, we can ignore the background radiation.\\ Second and third are the stability and confinement conditions. As we discussed earlier, stability condition tell us that the permissible region is the the lower region of the fig 1, and the confinement condition tell us the permissible region is the upper of the fig 2. Then the accepted region is the region between fig 1 and fig 2. Although entire points of this region are acceptable but we would like to bias our apparatus near the fig 2. Since we must prepare the system very sensitive, thus the electrons energy level must be near the potential barrier, so that by a reasonable change due to GW, the electrons get enough energy to pass the potential barrier. The stability condition and the confinement condition can be joined as a single condition. Actually, if we draw angular frequency in term of $\frac{2\omega_p^2}{\omega_c^2}$, we get the Fig-4 \cite{Krall}. For a given argument, we have two angular velocities: fast mode $\omega_+$ and the slow mode $\omega_-$. These two modes will be equal at the $\frac{2\omega_p^2}{\omega_c^2}=1$ (Brillouin flow).

\begin{figure}[tb]
  \centerline{\epsfxsize=4.5in \epsfysize=3.5in{\epsffile{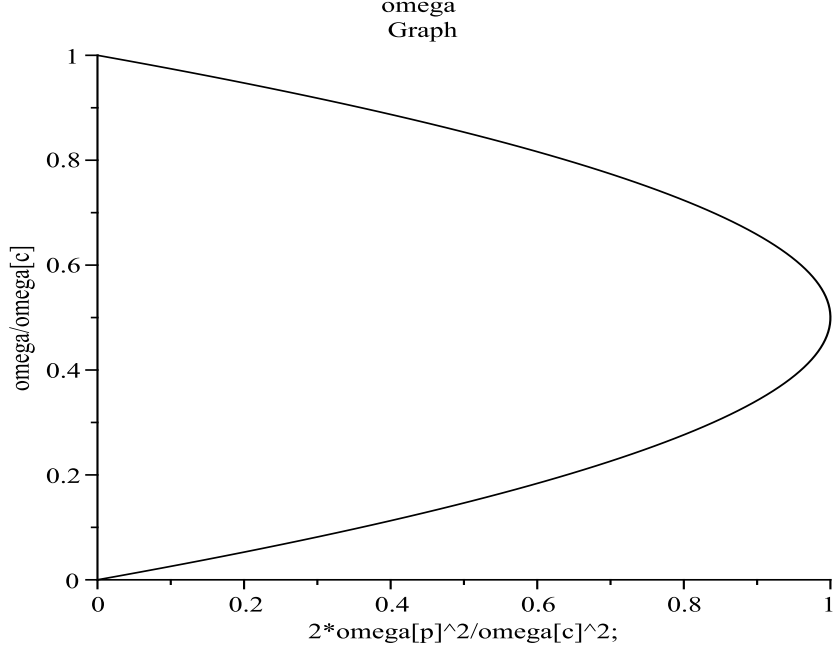}}}
    \caption{"Rigid-rotor" equilibrium angular velocity for an unneutralized uniform density cloud of electrons in a uniform magnetic field showing the fast, or axis-encircling, mode (upper branch) and the slow, or drift (lower branch) as a function of charge density \cite{Krall}.}\label{fgt}
\end{figure}

Now, the stability condition is equal to select one point of the Fig-4. When the electron have the maximum velocity then the confinement in a sensitive situation is achieved. Thus we must bias the electrons in the fast mode and in a little argument $\frac{2\omega_p^2}{\omega_c^2}$. As we can see from Fig-4,  $\omega$ will be maximum when $\frac{2\omega_p^2}{\omega_c^2}$ is near the zero. The amount of $\frac{2\omega_p^2}{\omega_c^2}(=\frac{8\pi n m c^2}{B^2})$ will be low if the $B$ is very high and $n$ is very low. Thus in constructing such a system, we select the maximum accessible value for  $B$  and the minimum feasible value for $n$. Being the $\omega$ maximum value, has an advantage. For a higher value of the electron velocity, we can sure the energy level is far from the ground state of the potential well, and then our classical calculation is reliable.
Now, we discuss about the wave numbers and the time frequency (not the rotational frequency) of the system. In general the wave numbers can be determined from the system's dimensions. The wave number along the $\phi$ is $k_\phi=\frac{2\pi}{2\pi}n_\phi=n_\phi$. If we concentrate on the low $n$ mode, then $k_\phi=1$.  Longitudinal wave number is equal to $k_z=\frac{2\pi}{l}n_z$, where $l$ is the length of the pipe. Having longitudinal wave number, the angular frequency can be obtained from the following dispersion relation:
\b (ka)^2=\frac{\omega^2 a^2}{c^2}-\frac{p_{n\nu}^2}{1-\omega_p^2/\omega^2},\e
where $a$ is radius of the pipe and $p_{n\nu}$ is the $\nu$th root of the $n$th-order Bessel's function of the first kind \cite{Krall}. \\
So far, we have obtained the wave numbers and frequency in term of dimensions of the system. Now, according to what was said earlier, we choose for $n$ a very low laboratory value such as $n=10^{10}{cm}^{-3}$ and for $B$ a very big laboratory value such as $B=10^3 G$. Then by adjusting the two far end potentials we confine the electrons in the pipe. To avoid the unwanted factors in ultimately bad situation be not able to pass electrons throw the potential barrier, we choose electron energy level for example $0.9$ of the potential height. Thus we must choose the following potential difference:

\b \triangle\phi=\frac{\frac{1}{2}m a \omega_r^2}{0.9 e}=10^{-7}_{volt}.\e

Now if the GW supplies the rest of the potential barrier, i.e, $0.1$ then some electron may go outside the rings and can be detected by a electron detector. Fig 5 show the perturbed energy in term of GW strength.

\begin{figure}[tb]
  \centerline{\epsfxsize=4.5in \epsfysize=3.5in{\epsffile{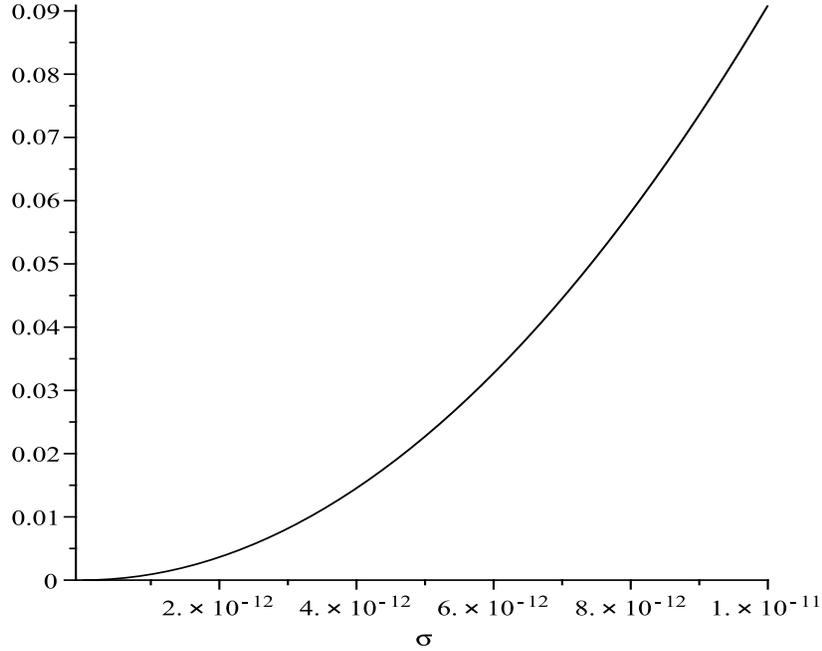}}}
    \caption{Ratio of perturbed electron energy to potential barrier height in term of GW strength. This ratio are obtained for the radius $a=10 cm$ and the length $ l=200 cm$ }\label{fgt}
\end{figure}

As we can see from the fig 5, this happen for the value $\sigma=10^{-11}$. In fact our proposed apparatus is successful detection device for such GW values.

\vspace{0.5cm} \noindent {\bf{Acknowledgements}}: Two of authors
(K. Mehdizadeh and O. Jalili) would
like to convey their gratitude to M. Jalili dean of Research Center
of Azad University at Amol. The authors also would like to express their thanks to Dr Sh. Rouhni and Dr M. Saravi for their comments and useful discussions.
 

\vspace{0.5cm}


\begin{thebibliography}{a}
\addcontentsline{toc}{chapter}{Bibliographie}

\bibitem{jalili}O. Jalili, M.V.Takook, Sh. Rouhani: Gravitational wave detection by bounded cold electronic plasma in a long pipe. International Journal of Theoretical Physics: Volume 49, Issue 1 (2010), Page 84.
\bibitem{Weber}J. Weber, General relativity and gravitational waves, Interscience, New York, 1961

\bibitem{david}Ronald C. Davidson: Theory of Nonneutral Plasmas, W.A. BENJAMIN, INC (1974)

\bibitem{Elli2} G F R Ellis : "Relativistic Cosmology" in Cargese Lectures in Physics, vol VI, ed. E. Schatzmann (Gordon and Breach, 1973).
\bibitem{Krall}Nicholas  A.  Krall , Alvin W. Trivelpiece, McGraw-Hill,inc(1973); Principle of plasma physics




\end{thebibliography}
\end{document}